\documentclass[12pt,letterpaper,english,titlepage]{article}
\usepackage[T1]{fontenc} 
\usepackage[utf8]{inputenc}
\usepackage{times}
\usepackage{geometry}
\usepackage{amsmath}
\usepackage{amsfonts}
\usepackage{amsthm}
\usepackage{amssymb}
\usepackage{color}
\RequirePackage{geometry}
\newcommand{\V}[1]{\mathbf{#1}} 

\usepackage[numbers,sort&compress]{natbib}
\title{Disentangling the Spatiotemporal Structure of Turbulence Using Multi-Spacecraft Data}
\author{J.~M.~TenBarge \\ 
Princeton University\\
Tel: 609-243-2603, Email: tenbarge@princeton.edu\vspace{1cm}\\
\underline{Co-Authors}\\
O.~Alexandrova$^1$, S.~Boldyrev$^2$, F.~Califano$^3$, S.~S.~Cerri$^4$, C.~H.~K.~Chen$^5$, \\
G.~G.~Howes$^6$, T.~Horbury$^7$, P.~A.~Isenberg$^8$, H.~Ji$^{9,10}$, K.~G.~Klein$^{11}$,\\ C.~Krafft$^{12,13}$, M.~Kunz$^{9,10}$, N.~F.~Loureiro$^{14}$, A.~Mallet$^{15}$, B.~A.~Maruca$^{16}$, \\
W.~H.~Matthaeus$^{16}$, R.~Meyrand$^{12,15}$, E.~Quataert$^{17}$, J.~C.~Perez$^{18}$,\\ O.~W.~Roberts$^{19}$, F.~Sahraoui$^{12}$, C.~S.~Salem$^{15}$, A.~A.~Schekochihin$^{20}$, H.~Spence$^{8}$, \\
J.~Squire$^{21}$, D. Told$^{22}$, D.~Verscharen$^{8,23}$, R.~T.~Wicks$^{23}$}
\date{%
    {\footnotesize 
    $^1$LESIA, Observatoire de Paris,
    $^2$University of Wisconsin,
    $^3$Universit\`{a} di Pisa,
    $^4$Princeton University,
    $^5$Queen Mary University of London,
    $^6$University of Iowa,
    $^7$Imperial College London,
    $^8$University of New Hampshire,
    $^9$Princeton University,
    $^{10}$Princeton Plasma Physics Laboratory,
    $^{11}$University of Arizona,
    $^{12}$Laboratoire de Physique des Plasmas, \'{E}cole Polytechnique,
    $^{13}$Univerit\'{e} Paris-Sud,
    $^{14}$Massachusetts Institute of Technology,
    $^{15}$Space Sciences Laboratory, University of California, Berkeley,
    $^{16}$University of Delaware,
    $^{17}$University of California, Berkeley,
    $^{18}$Florida Institute of Technology,
    $^{19}$\"{O}sterreichische Akademie der Wissenschaften, Institut f\"{u}r Weltraumforschung
    $^{20}$University of Oxford,
    $^{21}$University of Otago,
    $^{22}$Max-Planck-Institut f\"{u}r Plasmaphysik, Garching,
    $^{23}$Mullard Space Science Laboratory, University College London\\[2ex]}    
    \today}

\begin{document}
\maketitle
\begin{center}
{\Large {\bf Disentangling the Spatiotemporal Structure of Turbulence Using Multi-Spacecraft Data}}\vspace{-.35cm}
\end{center}

\section{Introduction}\vspace{-.45cm}
Turbulence in a magnetized plasma is the primary mechanism responsible for transforming energy at large injection scales into small-scale motions, which eventually dissipate, heating the plasma or accelerating particles. Plasma turbulence is ubiquitous in the universe, and it is responsible for the transport of mass, momentum, and energy in such diverse systems as the solar corona and wind, pulsar magnetospheres, accretion discs surrounding compact objects,  the interstellar medium, planet formation, and laboratory fusion devices. Indeed, under one of the four high-level science goals in the 2013 NRC Heliophysics Decadal survey states, to "[d]iscover and characterize fundamental processes that occur both within the heliosphere and throughout the universe," plasma turbulence is identified as a ubiquitous phenomenon involved in the energization of heliospheric and other astrophysical plasmas. 

Developing a predictive understanding of plasma turbulence is critical in heliospheric and astrophysical plasmas, both of which are systems in which the microscopic physics of turbulence can significantly impact the macroscopic evolution. For example, long standing questions in heliophysics -- such as how the solar corona is heated to temperatures that are orders of magnitude above that of the photosphere, or how the solar wind is launched from the Sun -- remain unanswered after decades of research, because we lack a detailed understanding of how the energy of turbulent plasma flows and electromagnetic fields is converted into plasma heat, or some other form of particle energization.

The vast majority of the plasma systems in the universe are weakly collisional, necessitating the application of kinetic plasma physics to fully understand them. Yet, kinetic plasma turbulence is an inherently multi-scale and multi-process phenomenon, coupling the largest scales of a system to sub-electron kinetic scales via a cascade of energy and also generating reconnecting current layers, shocks, and a myriad of instabilities and waves. The broad range of scales and processes encompassed by kinetic plasma turbulence preclude our ability to analytically or numerically model these global systems; therefore, we must turn to laboratory studies of confined plasmas or \textit{in situ} observations of natural plasmas, such as the solar wind, to advance the field. The solar wind is humankind's best resource for studying the naturally occurring plasmas that permeate the universe, and it is often referred to as a natural laboratory for plasma physics \cite{Tu:1995}. Since launching our first major scientific spacecraft mission, Explorer 1, in 1958, we have made significant progress characterizing solar wind turbulence. Yet, due to the severe limitations imposed by single point measurements, we are unable to characterize sufficiently the spatial and temporal properties of the solar wind, leaving many fundamental questions about plasma turbulence unanswered.  \textbf{Therefore, the time has now come wherein making  significant additional progress to determine the dynamical nature of solar wind turbulence requires multi-spacecraft missions spanning a wide range of scales simultaneously.} A dedicated multi-spacecraft mission concurrently covering a wide range of scales in the solar wind would not only allow us to directly determine the spatial and temporal structure of kinetic plasma turbulence, but it would also mitigate the limitations that current multi-spacecraft missions face, such as non-ideal orbits for observing solar wind turbulence. Some of the fundamentally important questions that can only be addressed by \textit{in situ} multipoint measurements are discussed below.


\section{Outstanding Questions}\vspace{-.4cm}
\subsection{What is the Energy Distribution in Frequency-Wavevector Space?}\vspace{-.25cm}
Measurements of turbulence in frequency-wavevector space provide insight into the dynamics of the plasma. However, all single point \textit{in situ} measurements rely on Taylor's hypothesis  \citep{Taylor:1938}, which assumes that the plasma does not evolve in time as it is convected past the spacecraft and establishes a direct connection between the frequency measured in the spacecraft-frame, $\omega_{sc}$, and the wavevector, $\V{k}$, of the fluctuations.  In the near-earth solar wind, the solar wind velocity is mostly radial and typically super-Alfv\'{e}nic, $v_{sw} \gg v_A$. Thus, observers adopt  Taylor's hypothesis by assuming that $|\omega| \ll |\V{k} \cdot \V{v}_{sw}|$, where $\omega$ is the plasma-frame frequency, so that the spacecraft-frame frequency fluctuations are interpreted to be related directly to the wavevector of the spatial fluctuations in the plasma frame, $\omega_{sc} \simeq \V{k}\cdot\V{v}_{sw}$ \citep{Matthaeus:1982,Perri:2010}. However, this central assumption has not been well tested in the solar wind, and it indeed fails when the solar wind speed is low compared to the Alfv\'{e}n speed or when  plasma-frame frequencies are high \citep{Howes:2014a,Klein:2014b}, as occurs with whistler or other dispersive fluctuations. \textbf{The only direct means of evaluating Taylor's hypothesis across a large range of scales is with a multipoint measurement in the solar wind to definitively determine if indeed $|\omega| \ll |\V{k} \cdot \V{v}_{sw}|$. }

Determining the plasma-frame frequency and wavevector distribution of energy is also fundamental for understanding the dynamics of turbulence. For instance, if on average $\omega(\V{k}) \ll \Omega_{cp}$, dissipation via resonant cyclotron damping is expected to be minimal, where  $\Omega_{cp}$ is the proton gyrofrequency. However, determining the plasma frame frequency requires fully resolving $\V{k}$. Even in cases wherein Taylor's hypothesis is well satisfied, single spacecraft measurements only provide access to the component of the wavevector along the solar wind flow direction. Existing multi-spacecraft missions such as Cluster have been employed to determine the plasma frame frequency, e.g., \citep{Sahraoui:2010b,Roberts:2015}; however, these studies have been limited to the approximately fifty, ten-minute intervals of Cluster data that satisfy the necessary conditions to apply multi-spacecraft techniques to the solar wind, namely that the four spacecraft are in a regular tetrahedron \citep{Narita:2010b,Sahraoui:2010a} and uncontaminated by foreshock particles backstreaming from earth's bowshock. To establish a statistical understanding of the energy distribution in frequency-wavevector space over a broad range of plasma parameters, a dedicated mission whose orbit is chosen to maximize time in the foreshock-free solar wind is necessary. \textbf{Thus, to unambiguously determine the  energy associated with the plasma frame frequency in the solar wind requires a dedicated suite of spacecraft, minimally four in a regular tetrahedron, or better yet, a swarm of craft covering a wide range of scales and angles, in the foreshock-free solar wind for extended periods of time.}

\subsection{What Dynamics Drive the Spectral Distribution of Turbulent Energy?}\label{sec:spectra}\vspace{-.25cm}
The distribution of energy in wavevector space is a core prediction of most plasma turbulence theories in use today. One popular example of turbulence theories which predict such a distributon is critical balance \citep{Goldreich:1995}, which assumes that the non-linear decorrelation time is equal to the linear propagation time, $\chi = \tau_{L}/\tau_{Nl} \sim 1$. This assumption has been modified and extended  to include the alignment of velocity and magnetic fluctuations with spatial scale, so called dynamical alignment \citep{Boldyrev:2005,Boldyrev:2006}, and most recently intermittency was incororporated in the model of refined critical balance \citep{Mallet:2015,Mallet:2017b}. At the heart of all of these models is the critical balance conjecture. Indeed, some research "suggests that critical balance... is the most robust and reliable of the physical principles underpinning theories of Alfv\'{e}nic turbulence" \citep{Mallet:2015} but theoretical alternatives to critical balance remain viable. Thus, measuring $\chi$ in the solar wind is the first step in testing the validity of all of these critical balance-based turbulence models, but the measurement requires resolving the components of the wavevector both parallel and perpendicular to the \textit{local in scale} magnetic field. Each of the turbulence models also predict different distributions of power in wavevector space, which again requires resolving the full wavevector.  Attempts have been made to resolve the wavevector using single spacecraft measurements, e.g., \cite{Horbury:2008,Podesta:2009a}, but these measurements require  $\V{v}_{sw}$ and $\V{B}$ to be aligned to determine $k_\parallel$, which is a rare occurrence. Therefore, single spacecraft tests of these models require combining several days to a month of data, while the auto-correlation time in the solar wind is of order one hour or less, likely mixing different plasma and turbulence conditions in the analysis. Cluster has been used to determine the full spectral anisotropy \cite{Roberts:2017}; however, the range of accessible scales were highly limited. \textbf{Thus, definitively determining the validity of each of these turbulence models or making progress in developing new models requires a multi-spacecraft mission, where all components of the wavevector can be measured simultaneously, rather than combining many single point datasets spanning turbulence with widely varying parameters.} 


\subsection{What is the Turbulent Cascade Rate?}\vspace{-.25cm}
The cascade of energy in a plasma can be directly measured using third order statistics \citep{Politano:1998}, and the cascade rate is related to proton heating in the solar wind \citep{Stawarz:2009}. Under a certain set of assumptions, Kolmogorov's third order law is the only exact, non-trivial turbulence result in hydrodynamics \citep{Frisch:1995}. A similar exact result exists for plasma turbulence under a more restrictive set of assumptions. However, the plasma turbulence cascade is anisotropic, and a single spacecraft can not resolve the anisotropy. Therefore, a multipoint measurement is necessary to properly measure the cascade. \textbf{A multipoint measurement spanning many scales can not only measure the anisotropy, it can measure the spatial gradients contained in the third order equation, directly accessing the primitive form of the third order law and for the first time, bypassing any assumptions about isotropy.}

\subsection{What is the Nature of Intermittency?}\vspace{-.25cm}
Intermittency, or patchiness in space and time, is an essential property of turbulence directly related to the cascade and dissipation of energy. The stationarity assumption made in single point spacecraft observations necessarily means it is impossible to disentangle the spatial or temporal nature of intermittent fluctuations, and a causal connection between the structures and observed local heating \citep{Osman:2011,Osman:2011a} is equally difficult \citep{Borovsky:2011}. Single point measurements also cannot provide information about the 3D structure and nature of the intermittency. Turbulence models like refined critical balance make predictions about the nature and scaling of intermittent structures \citep{Mallet:2017b}; however, testing the predictions requires measuring coherence lengths parallel and perpendicular to the local magnetic field. Such a test is not possible with a singe spacecraft without combining many epochs of \textit{in situ} data with different plasma conditions. Single point \textit{in situ} observations have also found that there is a transition "from multifractal intermittent turbulence in the inertial range to non-Gaussian mono-scaling in the dissipation range" \citep{Kiyani:2009}, which is not a phenomenon observed nor predicted in hydrodynamic or plasma turbulence. \textbf{A swarm of spacecraft observing the solar wind at a large range of scales could directly address open questions regarding the nature and origin of intermittency.}

\subsection{What is the Spatial Distribution of Turbulent Fluctuations?}\vspace{-.25cm}
Many processes that operate in the solar wind locally generate structures or waves, and turbulence itself is inherently intermittent; however, single spacecraft measurements cannot disentangle the causality, evolution, or distribution of the processes, because the implicit assumption is that the plasma is stationary over the period of the measurement. 



One example of a process that is impossible to characterize fully using single spacecraft observations is \textbf{instabilities and wave generation.} Wave modes such as the mirror, whistler, and ion cyclotron modes are routinely observed to constitute a small fraction of the solar wind, e.g., \citep{Podesta:2011a,Klein:2014a,Lacombe:2014}; however, most of these modes are not expected to exist in the solar wind. Many instabilities expected to operate in the solar wind saturate by generating such wave modes, but a causal connection with regions of unstable plasma and the observed modes is difficult with single spacecraft measurements, although attempts have been made \citep{Gary:2016}. Also, the sub-dominant modes likely generated by instabilities are difficult to resolve, because they are masked by the more energetic Alfv\'{e}nic turbulence. These instability associated modes also provide a non-local method to transfer energy from large to small scales, bypassing the turbulent cascade \citep{Kunz:2014,Riquelme:2015}.  Despite being sub-dominant, the modes generated by instabilities are high amplitude and can efficiently scatter particles, leading to fluid-like behavior, including viscous dissipation, even in weakly collisional plasmas \citep{Kunz:2014,Verscharen:2016,Riquelme:2016}. The same instabilities can also lead to complete disruption of Alv\'{e}nic fluctuations  \citep{Squire:2016,Squire:2017}, which could potentially terminate turbulence cascade entirely. \textbf{Therefore, it is fundamentally important to establish how frequently these modes are present and the causal connection between the modes and progenitor instabilities, which is only possible with a multipoint observation.}

\section{Conclusion}\vspace{-.4cm}

In summary, plasma turbulence plays a fundamental role in the transport of energy, mass, and momentum in the universe. Progress in understanding turbulence will benefit many areas, including fusion confinement, interpreting astrophysical observations, space weather, and the coronal heating problem; however, we have reached a point wherein progress is inhibited by the paucity of multipoint \textit{in situ} solar wind measurements. \textbf{Bringing closure to the spatiotemporal structure of turbulence will be transformative for the field, and it can only be fully addressed with multipoint measurements, requiring a swarm of spacecraft spanning a wide range of scales.} 

\vspace{-1.5cm}
\bibliographystyle{apsrev}
\bibliography{abbrev2,all}

\begin{thebibliography}{36}
\expandafter\ifx\csname natexlab\endcsname\relax\def\natexlab#1{#1}\fi
\expandafter\ifx\csname bibnamefont\endcsname\relax
  \def\bibnamefont#1{#1}\fi
\expandafter\ifx\csname bibfnamefont\endcsname\relax
  \def\bibfnamefont#1{#1}\fi
\expandafter\ifx\csname citenamefont\endcsname\relax
  \def\citenamefont#1{#1}\fi
\expandafter\ifx\csname url\endcsname\relax
  \def\url#1{\texttt{#1}}\fi
\expandafter\ifx\csname urlprefix\endcsname\relax\def\urlprefix{URL }\fi
\providecommand{\bibinfo}[2]{#2}
\providecommand{\eprint}[2][]{}

\bibitem[{\citenamefont{{Tu} and {Marsch}}(1995)}]{Tu:1995}
\bibinfo{author}{\bibfnamefont{C.-Y.} \bibnamefont{{Tu}}} \bibnamefont{and}
  \bibinfo{author}{\bibfnamefont{E.}~\bibnamefont{{Marsch}}},
  \bibinfo{journal}{Space Science Reviews} \textbf{\bibinfo{volume}{73}},
  \bibinfo{pages}{1} (\bibinfo{year}{1995}).

\bibitem[{\citenamefont{{Taylor}}(1938)}]{Taylor:1938}
\bibinfo{author}{\bibfnamefont{G.~I.} \bibnamefont{{Taylor}}},
  \bibinfo{journal}{{Proc. Roy. Soc. A}} \textbf{\bibinfo{volume}{164}},
  \bibinfo{pages}{476} (\bibinfo{year}{1938}).

\bibitem[{\citenamefont{{Matthaeus} and {Goldstein}}(1982)}]{Matthaeus:1982}
\bibinfo{author}{\bibfnamefont{W.~H.} \bibnamefont{{Matthaeus}}}
  \bibnamefont{and} \bibinfo{author}{\bibfnamefont{M.~L.}
  \bibnamefont{{Goldstein}}}, \bibinfo{journal}{J.~Geophys.~Res.}
  \textbf{\bibinfo{volume}{87}}, \bibinfo{pages}{10347} (\bibinfo{year}{1982}).

\bibitem[{\citenamefont{{Perri} and {Balogh}}(2010)}]{Perri:2010}
\bibinfo{author}{\bibfnamefont{S.}~\bibnamefont{{Perri}}} \bibnamefont{and}
  \bibinfo{author}{\bibfnamefont{A.}~\bibnamefont{{Balogh}}},
  \bibinfo{journal}{Astrophys.~J.} \textbf{\bibinfo{volume}{714}},
  \bibinfo{pages}{937} (\bibinfo{year}{2010}).

\bibitem[{\citenamefont{{Howes} et~al.}(2014)\citenamefont{{Howes}, {Klein},
  and {TenBarge}}}]{Howes:2014a}
\bibinfo{author}{\bibfnamefont{G.~G.} \bibnamefont{{Howes}}},
  \bibinfo{author}{\bibfnamefont{K.~G.} \bibnamefont{{Klein}}},
  \bibnamefont{and} \bibinfo{author}{\bibfnamefont{J.~M.}
  \bibnamefont{{TenBarge}}}, \bibinfo{journal}{Astrophys.~J.}
  \textbf{\bibinfo{volume}{789}}, \bibinfo{eid}{106} (\bibinfo{year}{2014}).

\bibitem[{\citenamefont{{Klein}
  et~al.}(2014{\natexlab{a}})\citenamefont{{Klein}, {Howes}, and
  {TenBarge}}}]{Klein:2014b}
\bibinfo{author}{\bibfnamefont{K.~G.} \bibnamefont{{Klein}}},
  \bibinfo{author}{\bibfnamefont{G.~G.} \bibnamefont{{Howes}}},
  \bibnamefont{and} \bibinfo{author}{\bibfnamefont{J.~M.}
  \bibnamefont{{TenBarge}}}, \bibinfo{journal}{Astrophys.~J.~Lett.}
  \textbf{\bibinfo{volume}{790}}, \bibinfo{eid}{L20}
  (\bibinfo{year}{2014}{\natexlab{a}}).

\bibitem[{\citenamefont{{Sahraoui}
  et~al.}(2010{\natexlab{a}})\citenamefont{{Sahraoui}, {Goldstein}, {Belmont},
  {Canu}, and {Rezeau}}}]{Sahraoui:2010b}
\bibinfo{author}{\bibfnamefont{F.}~\bibnamefont{{Sahraoui}}},
  \bibinfo{author}{\bibfnamefont{M.~L.} \bibnamefont{{Goldstein}}},
  \bibinfo{author}{\bibfnamefont{G.}~\bibnamefont{{Belmont}}},
  \bibinfo{author}{\bibfnamefont{P.}~\bibnamefont{{Canu}}}, \bibnamefont{and}
  \bibinfo{author}{\bibfnamefont{L.}~\bibnamefont{{Rezeau}}},
  \bibinfo{journal}{Phys.~Rev.~Lett.} \textbf{\bibinfo{volume}{105}},
  \bibinfo{pages}{131101} (\bibinfo{year}{2010}{\natexlab{a}}).

\bibitem[{\citenamefont{{Roberts} et~al.}(2015)\citenamefont{{Roberts}, {Li},
  and {Jeska}}}]{Roberts:2015}
\bibinfo{author}{\bibfnamefont{O.~W.} \bibnamefont{{Roberts}}},
  \bibinfo{author}{\bibfnamefont{X.}~\bibnamefont{{Li}}}, \bibnamefont{and}
  \bibinfo{author}{\bibfnamefont{L.}~\bibnamefont{{Jeska}}},
  \bibinfo{journal}{Astrophys.~J.} \textbf{\bibinfo{volume}{802}},
  \bibinfo{eid}{2} (\bibinfo{year}{2015}).

\bibitem[{\citenamefont{{Narita} et~al.}(2010)\citenamefont{{Narita},
  {Glassmeier}, and {Motschmann}}}]{Narita:2010b}
\bibinfo{author}{\bibfnamefont{Y.}~\bibnamefont{{Narita}}},
  \bibinfo{author}{\bibfnamefont{K.-H.} \bibnamefont{{Glassmeier}}},
  \bibnamefont{and}
  \bibinfo{author}{\bibfnamefont{U.}~\bibnamefont{{Motschmann}}},
  \bibinfo{journal}{Nonlin.~Proc.~Geophys.} \textbf{\bibinfo{volume}{17}},
  \bibinfo{pages}{383} (\bibinfo{year}{2010}).

\bibitem[{\citenamefont{{Sahraoui}
  et~al.}(2010{\natexlab{b}})\citenamefont{{Sahraoui}, {Belmont}, {Goldstein},
  and {Rezeau}}}]{Sahraoui:2010a}
\bibinfo{author}{\bibfnamefont{F.}~\bibnamefont{{Sahraoui}}},
  \bibinfo{author}{\bibfnamefont{G.}~\bibnamefont{{Belmont}}},
  \bibinfo{author}{\bibfnamefont{M.~L.} \bibnamefont{{Goldstein}}},
  \bibnamefont{and} \bibinfo{author}{\bibfnamefont{L.}~\bibnamefont{{Rezeau}}},
  \bibinfo{journal}{J.~Geophys.~Res.} \textbf{\bibinfo{volume}{115}},
  \bibinfo{pages}{4206} (\bibinfo{year}{2010}{\natexlab{b}}).

\bibitem[{\citenamefont{Goldreich and Sridhar}(1995)}]{Goldreich:1995}
\bibinfo{author}{\bibfnamefont{P.}~\bibnamefont{Goldreich}} \bibnamefont{and}
  \bibinfo{author}{\bibfnamefont{S.}~\bibnamefont{Sridhar}},
  \bibinfo{journal}{Astrophys.~J.} \textbf{\bibinfo{volume}{438}},
  \bibinfo{pages}{763} (\bibinfo{year}{1995}).

\bibitem[{\citenamefont{{Boldyrev}}(2005)}]{Boldyrev:2005}
\bibinfo{author}{\bibfnamefont{S.}~\bibnamefont{{Boldyrev}}},
  \bibinfo{journal}{Astrophys.~J.~Lett.} \textbf{\bibinfo{volume}{626}},
  \bibinfo{pages}{L37} (\bibinfo{year}{2005}).

\bibitem[{\citenamefont{{Boldyrev}}(2006)}]{Boldyrev:2006}
\bibinfo{author}{\bibfnamefont{S.}~\bibnamefont{{Boldyrev}}},
  \bibinfo{journal}{Phys.~Rev.~Lett.} \textbf{\bibinfo{volume}{96}},
  \bibinfo{pages}{115002} (\bibinfo{year}{2006}).

\bibitem[{\citenamefont{{Mallet} et~al.}(2015)\citenamefont{{Mallet},
  {Schekochihin}, and {Chandran}}}]{Mallet:2015}
\bibinfo{author}{\bibfnamefont{A.}~\bibnamefont{{Mallet}}},
  \bibinfo{author}{\bibfnamefont{A.~A.} \bibnamefont{{Schekochihin}}},
  \bibnamefont{and} \bibinfo{author}{\bibfnamefont{B.~D.~G.}
  \bibnamefont{{Chandran}}}, \bibinfo{journal}{Mon.~Not.~Roy.~Astron.~Soc.}
  \textbf{\bibinfo{volume}{449}}, \bibinfo{pages}{L77} (\bibinfo{year}{2015}).

\bibitem[{\citenamefont{{Mallet} and {Schekochihin}}(2017)}]{Mallet:2017b}
\bibinfo{author}{\bibfnamefont{A.}~\bibnamefont{{Mallet}}} \bibnamefont{and}
  \bibinfo{author}{\bibfnamefont{A.~A.} \bibnamefont{{Schekochihin}}},
  \bibinfo{journal}{Mon.~Not.~Roy.~Astron.~Soc.}
  \textbf{\bibinfo{volume}{466}}, \bibinfo{pages}{3918} (\bibinfo{year}{2017}).

\bibitem[{\citenamefont{{Horbury} et~al.}(2008)\citenamefont{{Horbury}, Forman,
  and Oughton}}]{Horbury:2008}
\bibinfo{author}{\bibfnamefont{T.~S.} \bibnamefont{{Horbury}}},
  \bibinfo{author}{\bibfnamefont{M.}~\bibnamefont{Forman}}, \bibnamefont{and}
  \bibinfo{author}{\bibfnamefont{S.}~\bibnamefont{Oughton}},
  \bibinfo{journal}{Phys.~Rev.~Lett.} \textbf{\bibinfo{volume}{101}},
  \bibinfo{pages}{175005} (\bibinfo{year}{2008}).

\bibitem[{\citenamefont{{Podesta}}(2009)}]{Podesta:2009a}
\bibinfo{author}{\bibfnamefont{J.~J.} \bibnamefont{{Podesta}}},
  \bibinfo{journal}{Astrophys.~J.} \textbf{\bibinfo{volume}{698}},
  \bibinfo{pages}{986} (\bibinfo{year}{2009}).

\bibitem[{\citenamefont{{Roberts} et~al.}(2017)\citenamefont{{Roberts},
  {Narita}, and {Escoubet}}}]{Roberts:2017}
\bibinfo{author}{\bibfnamefont{O.~W.} \bibnamefont{{Roberts}}},
  \bibinfo{author}{\bibfnamefont{Y.}~\bibnamefont{{Narita}}}, \bibnamefont{and}
  \bibinfo{author}{\bibfnamefont{C.~P.} \bibnamefont{{Escoubet}}},
  \bibinfo{journal}{Astrophys.~J.~Lett.} \textbf{\bibinfo{volume}{851}},
  \bibinfo{eid}{L11} (\bibinfo{year}{2017}).

\bibitem[{\citenamefont{{Politano} and {Pouquet}}(1998)}]{Politano:1998}
\bibinfo{author}{\bibfnamefont{H.}~\bibnamefont{{Politano}}} \bibnamefont{and}
  \bibinfo{author}{\bibfnamefont{A.}~\bibnamefont{{Pouquet}}},
  \bibinfo{journal}{Phys.~Rev.~E} \textbf{\bibinfo{volume}{57}},
  \bibinfo{pages}{R21} (\bibinfo{year}{1998}).

\bibitem[{\citenamefont{{Stawarz} et~al.}(2009)\citenamefont{{Stawarz},
  {Smith}, {Vasquez}, {Forman}, and {MacBride}}}]{Stawarz:2009}
\bibinfo{author}{\bibfnamefont{J.~E.} \bibnamefont{{Stawarz}}},
  \bibinfo{author}{\bibfnamefont{C.~W.} \bibnamefont{{Smith}}},
  \bibinfo{author}{\bibfnamefont{B.~J.} \bibnamefont{{Vasquez}}},
  \bibinfo{author}{\bibfnamefont{M.~A.} \bibnamefont{{Forman}}},
  \bibnamefont{and} \bibinfo{author}{\bibfnamefont{B.~T.}
  \bibnamefont{{MacBride}}}, \bibinfo{journal}{Astrophys.~J.}
  \textbf{\bibinfo{volume}{697}}, \bibinfo{pages}{1119} (\bibinfo{year}{2009}).

\bibitem[{\citenamefont{{Frisch}}(1995)}]{Frisch:1995}
\bibinfo{author}{\bibfnamefont{U.}~\bibnamefont{{Frisch}}},
  \emph{\bibinfo{title}{{Turbulence. The legacy of A. N. Kolmogorov.}}}
  (\bibinfo{publisher}{Cambridge University Press}, \bibinfo{year}{1995}).

\bibitem[{\citenamefont{{Osman}
  et~al.}(2011{\natexlab{a}})\citenamefont{{Osman}, {Matthaeus}, {Greco}, and
  {Servidio}}}]{Osman:2011}
\bibinfo{author}{\bibfnamefont{K.~T.} \bibnamefont{{Osman}}},
  \bibinfo{author}{\bibfnamefont{W.~H.} \bibnamefont{{Matthaeus}}},
  \bibinfo{author}{\bibfnamefont{A.}~\bibnamefont{{Greco}}}, \bibnamefont{and}
  \bibinfo{author}{\bibfnamefont{S.}~\bibnamefont{{Servidio}}},
  \bibinfo{journal}{Astrophys.~J.~Lett.} \textbf{\bibinfo{volume}{727}},
  \bibinfo{eid}{L11} (\bibinfo{year}{2011}{\natexlab{a}}).

\bibitem[{\citenamefont{{Osman}
  et~al.}(2011{\natexlab{b}})\citenamefont{{Osman}, {Matthaeus}, {Wan}, and
  {Rappazzo}}}]{Osman:2011a}
\bibinfo{author}{\bibfnamefont{K.~T.} \bibnamefont{{Osman}}},
  \bibinfo{author}{\bibfnamefont{W.~H.} \bibnamefont{{Matthaeus}}},
  \bibinfo{author}{\bibfnamefont{M.}~\bibnamefont{{Wan}}}, \bibnamefont{and}
  \bibinfo{author}{\bibfnamefont{A.~F.} \bibnamefont{{Rappazzo}}},
  \bibinfo{journal}{ArXiv e-prints} \textbf{\bibinfo{volume}{108}},
  \bibinfo{eid}{261102} (\bibinfo{year}{2011}{\natexlab{b}}).

\bibitem[{\citenamefont{{Borovsky} and {Denton}}(2011)}]{Borovsky:2011}
\bibinfo{author}{\bibfnamefont{J.~E.} \bibnamefont{{Borovsky}}}
  \bibnamefont{and} \bibinfo{author}{\bibfnamefont{M.~H.}
  \bibnamefont{{Denton}}}, \bibinfo{journal}{Astrophys.~J.~Lett.}
  \textbf{\bibinfo{volume}{739}}, \bibinfo{eid}{L61} (\bibinfo{year}{2011}).

\bibitem[{\citenamefont{{Kiyani} et~al.}(2009)\citenamefont{{Kiyani},
  {Chapman}, {Khotyaintsev}, {Dunlop}, and {Sahraoui}}}]{Kiyani:2009}
\bibinfo{author}{\bibfnamefont{K.~H.} \bibnamefont{{Kiyani}}},
  \bibinfo{author}{\bibfnamefont{S.~C.} \bibnamefont{{Chapman}}},
  \bibinfo{author}{\bibfnamefont{Y.~V.} \bibnamefont{{Khotyaintsev}}},
  \bibinfo{author}{\bibfnamefont{M.~W.} \bibnamefont{{Dunlop}}},
  \bibnamefont{and}
  \bibinfo{author}{\bibfnamefont{F.}~\bibnamefont{{Sahraoui}}},
  \bibinfo{journal}{Physical Review Letters} \textbf{\bibinfo{volume}{103}},
  \bibinfo{pages}{075006} (\bibinfo{year}{2009}).

\bibitem[{\citenamefont{{Podesta} and {Gary}}(2011)}]{Podesta:2011a}
\bibinfo{author}{\bibfnamefont{J.~J.} \bibnamefont{{Podesta}}}
  \bibnamefont{and} \bibinfo{author}{\bibfnamefont{S.~P.}
  \bibnamefont{{Gary}}}, \bibinfo{journal}{Astrophys.~J.}
  \textbf{\bibinfo{volume}{734}}, \bibinfo{pages}{15} (\bibinfo{year}{2011}).

\bibitem[{\citenamefont{{Klein}
  et~al.}(2014{\natexlab{b}})\citenamefont{{Klein}, {Howes}, {TenBarge}, and
  {Podesta}}}]{Klein:2014a}
\bibinfo{author}{\bibfnamefont{K.~G.} \bibnamefont{{Klein}}},
  \bibinfo{author}{\bibfnamefont{G.~G.} \bibnamefont{{Howes}}},
  \bibinfo{author}{\bibfnamefont{J.~M.} \bibnamefont{{TenBarge}}},
  \bibnamefont{and} \bibinfo{author}{\bibfnamefont{J.~J.}
  \bibnamefont{{Podesta}}}, \bibinfo{journal}{Astrophys.~J.}
  \textbf{\bibinfo{volume}{785}}, \bibinfo{eid}{138}
  (\bibinfo{year}{2014}{\natexlab{b}}).

  \bibitem[{\citenamefont{{Lacombe} et~al.}(2014)\citenamefont{{Lacombe},
  {Alexandrova}, {Matteini}, {Santol{\'{\i}}k}, {Cornilleau-Wehrlin},
  {Mangeney}, {de Conchy}, and {Maksimovic}}}]{Lacombe:2014}
\bibinfo{author}{\bibfnamefont{C.}~\bibnamefont{{Lacombe}}},
  \bibinfo{author}{\bibfnamefont{O.}~\bibnamefont{{Alexandrova}}},
  \bibinfo{author}{\bibfnamefont{L.}~\bibnamefont{{Matteini}}},
  \bibinfo{author}{\bibfnamefont{O.}~\bibnamefont{{Santol{\'{\i}}k}}},
  \bibinfo{author}{\bibfnamefont{N.}~\bibnamefont{{Cornilleau-Wehrlin}}},
  \bibinfo{author}{\bibfnamefont{A.}~\bibnamefont{{Mangeney}}},
  \bibinfo{author}{\bibfnamefont{Y.}~\bibnamefont{{de Conchy}}},
  \bibnamefont{and}
  \bibinfo{author}{\bibfnamefont{M.}~\bibnamefont{{Maksimovic}}},
  \bibinfo{journal}{Astrophys.~J.} \textbf{\bibinfo{volume}{796}},
  \bibinfo{eid}{5} (\bibinfo{year}{2014}).

\bibitem[{\citenamefont{{Gary} et~al.}(2016)\citenamefont{{Gary}, {Jian},
  {Broiles}, {Stevens}, {Podesta}, and {Kasper}}}]{Gary:2016}
\bibinfo{author}{\bibfnamefont{S.~P.} \bibnamefont{{Gary}}},
  \bibinfo{author}{\bibfnamefont{L.~K.} \bibnamefont{{Jian}}},
  \bibinfo{author}{\bibfnamefont{T.~W.} \bibnamefont{{Broiles}}},
  \bibinfo{author}{\bibfnamefont{M.~L.} \bibnamefont{{Stevens}}},
  \bibinfo{author}{\bibfnamefont{J.~J.} \bibnamefont{{Podesta}}},
  \bibnamefont{and} \bibinfo{author}{\bibfnamefont{J.~C.}
  \bibnamefont{{Kasper}}}, \bibinfo{journal}{J.~Geophys.~Res.}
  \textbf{\bibinfo{volume}{121}}, \bibinfo{pages}{30} (\bibinfo{year}{2016}).

\bibitem[{\citenamefont{{Kunz} et~al.}(2014)\citenamefont{{Kunz},
  {Schekochihin}, and {Stone}}}]{Kunz:2014}
\bibinfo{author}{\bibfnamefont{M.~W.} \bibnamefont{{Kunz}}},
  \bibinfo{author}{\bibfnamefont{A.~A.} \bibnamefont{{Schekochihin}}},
  \bibnamefont{and} \bibinfo{author}{\bibfnamefont{J.~M.}
  \bibnamefont{{Stone}}}, \bibinfo{journal}{Phys.~Rev.~Lett.}
  \textbf{\bibinfo{volume}{112}}, \bibinfo{pages}{205003}
  (\bibinfo{year}{2014}).

\bibitem[{\citenamefont{{Riquelme} et~al.}(2015)\citenamefont{{Riquelme},
  {Quataert}, and {Verscharen}}}]{Riquelme:2015}
\bibinfo{author}{\bibfnamefont{M.~A.} \bibnamefont{{Riquelme}}},
  \bibinfo{author}{\bibfnamefont{E.}~\bibnamefont{{Quataert}}},
  \bibnamefont{and}
  \bibinfo{author}{\bibfnamefont{D.}~\bibnamefont{{Verscharen}}},
  \bibinfo{journal}{Astrophys.~J.} \textbf{\bibinfo{volume}{800}},
  \bibinfo{eid}{27} (\bibinfo{year}{2015}).

\bibitem[{\citenamefont{{Verscharen} et~al.}(2016)\citenamefont{{Verscharen},
  {Chandran}, {Klein}, and {Quataert}}}]{Verscharen:2016}
\bibinfo{author}{\bibfnamefont{D.}~\bibnamefont{{Verscharen}}},
  \bibinfo{author}{\bibfnamefont{B.~D.~G.} \bibnamefont{{Chandran}}},
  \bibinfo{author}{\bibfnamefont{K.~G.} \bibnamefont{{Klein}}},
  \bibnamefont{and}
  \bibinfo{author}{\bibfnamefont{E.}~\bibnamefont{{Quataert}}},
  \bibinfo{journal}{Astrophys.~J.} \textbf{\bibinfo{volume}{831}},
  \bibinfo{eid}{128} (\bibinfo{year}{2016}).

\bibitem[{\citenamefont{{Riquelme} et~al.}(2016)\citenamefont{{Riquelme},
  {Quataert}, and {Verscharen}}}]{Riquelme:2016}
\bibinfo{author}{\bibfnamefont{M.~A.} \bibnamefont{{Riquelme}}},
  \bibinfo{author}{\bibfnamefont{E.}~\bibnamefont{{Quataert}}},
  \bibnamefont{and}
  \bibinfo{author}{\bibfnamefont{D.}~\bibnamefont{{Verscharen}}},
  \bibinfo{journal}{Astrophys.~J.} \textbf{\bibinfo{volume}{824}},
  \bibinfo{eid}{123} (\bibinfo{year}{2016}).

\bibitem[{\citenamefont{{Squire} et~al.}(2016)\citenamefont{{Squire},
  {Quataert}, and {Schekochihin}}}]{Squire:2016}
\bibinfo{author}{\bibfnamefont{J.}~\bibnamefont{{Squire}}},
  \bibinfo{author}{\bibfnamefont{E.}~\bibnamefont{{Quataert}}},
  \bibnamefont{and} \bibinfo{author}{\bibfnamefont{A.~A.}
  \bibnamefont{{Schekochihin}}}, \bibinfo{journal}{Astrophys.~J.~Lett.}
  \textbf{\bibinfo{volume}{830}}, \bibinfo{eid}{L25} (\bibinfo{year}{2016}).

\bibitem[{\citenamefont{{Squire} et~al.}(2017)\citenamefont{{Squire}, {Kunz},
  {Quataert}, and {Schekochihin}}}]{Squire:2017}
\bibinfo{author}{\bibfnamefont{J.}~\bibnamefont{{Squire}}},
  \bibinfo{author}{\bibfnamefont{M.~W.} \bibnamefont{{Kunz}}},
  \bibinfo{author}{\bibfnamefont{E.}~\bibnamefont{{Quataert}}},
  \bibnamefont{and} \bibinfo{author}{\bibfnamefont{A.~A.}
  \bibnamefont{{Schekochihin}}}, \bibinfo{journal}{Phys.~Rev.~Lett.}
  \textbf{\bibinfo{volume}{119}}, \bibinfo{eid}{155101} (\bibinfo{year}{2017}).

\end{thebibliography}

\end{document}